\documentclass[10pt]{article}
\usepackage[margin=3cm]{geometry}
\usepackage{polski}
\usepackage[utf8]{inputenc}
\usepackage{graphicx}
\usepackage{amssymb}
\usepackage{amsmath} 
\usepackage{dsfont}

\usepackage{authblk}

\usepackage[backend=biber, 
sorting=none,
style=numeric-comp]{biblatex}
\usepackage{subcaption}
\usepackage{ stmaryrd }

\newcommand{\1}{\hat{\mathds{1}}}
\newcommand{\q}{\hat{Q}}
\newcommand{\p}{\hat{P}}
\newcommand{\D}{\hat{D}}

\title{Quantum dynamics in Weyl-Heisenberg coherent states}
\author{Artur Miroszewski \\ \emph{artur.miroszewski@ncbj.gov.pl}}
\date{}
\affil{\textit{National Centre for Nuclear Research}\\ \textit{Ludwika Pasteura 7, 02-093, Warsaw, Poland}}

\begin{document}

\maketitle

\begin{abstract}
The article explores a new formalism for describing motion in quantum mechanics. The construction is based on generalized coherent states with evolving fiducial vector. Weyl-Heisenberg coherent states are utilised to split quantum systems into `classical' and `quantum' degrees of freedom. The decomposition is found to be equivalent to quantum mechanics perceived from a semi-classical frame. The split allows for introduction of a new definition of classical state and is a convenient starting point for approximate analysis of quantum dynamics. An example of a meta-stable state is given as a practical illustration of the introduced concepts.
\end{abstract}
\section{Introduction}

Coherent states have occupied physicists and mathematicians for almost a century. First introduced in 1926 by Edwin Schr\"odinger \cite{Schrodinger:1926} in their standard formulation and studied by John von Neumann \cite{vonNeumann:1932} from the phase space perspective they were forgotten until beginning of the 1960s. Recognizing their usefulness in the subject of atomic optics \cite{Glauber:1963I, Glauber:1963II}, introduction of the concept of generalized coherent states \cite{Klauder:1968, Klauder:1963I, Klauder:1963II} and their connection to group theory \cite{Perelomov:1972} resulted in unflagging interest in coherent states until today. Their success in the physical sciences can be seen from the perspective of the amount of fields which employed coherent states as an effective tool. Among others, superfluidity \cite{Perelomov:1977}, superradiance \cite{Wang:1973, Hepp:1973}, quantum electrodynamics \cite{Chung:1965, Kibble:1968I, Kibble:1968II}, solitons  \cite{Shapiro:1976, Cahill:1974, Vinciarelli:1975}, statistical physics and semiclassical limits \cite{Lieb:1973, Simon:1980}, scattering processes \cite{Carruthers:1965} and recently quantum cosmology \cite{Bergeron:2014, Malkiewicz:2018, Bergeron:2017}. Along with solving the dynamics of some observables in particular physical systems exactly, coherent states gave rise to the new quantisation methods, connecting real functions of the phase space variables to the self-adjoint operators \cite{Bergeron:2017II, Klauder:2015, Bergeron:2001, Kirillov:1976, Kostant:1970, Kostant1970I, Berezin:1975}. For more comprehensive reviews of the field of coherent states see \cite{Gazeau:2009, Perelomov:1986}.\\
The background motivation for this paper originates in the stunning property of coherent states, being able to connect the realms of classical and quantum mechanics. On one hand one can go from the classical systems to quantum ones using coherent states quantisation methods. Those methods provide a general and robust procedure based on the symmetries of the background phase space, allowing one not only to perform the standard canonical quantisation, but also to quantise the systems on non-trivial phase spaces, like half-plane or sphere \cite{Gazeau:2009}. Many genuine ambiguities of the quantisation methods are reduced to the recognition of the symmetry group of the problem and a single choice of the fiducial wavefunction, which seeds the whole procedure.\\
On the other hand one can reduce the quantum action by narrowing down the states available in the dynamics
to coherent states \cite{Klauder:2015}. This approach allows one to project full quantum dynamics to the semi-classical phase space of coherent state’s parameters. The evolution of the parameters is governed by Hamilton-like equations, which are reduced to the standard Hamilton equations in the limit $\hbar \rightarrow 0$. The form of the Hamilton-like equations directly depends on the initial shape of the fiducial wavefunction, which, at least quantitatively, leads to the ambiguity of the dynamics. The origin of the ambiguity is simple; when treating the reduced system, one chooses the properties of the fiducial wavefunction arbitrarily and implicitly sets them fixed, while in full quantum mechanics the wavefunction changes as the system evolves.\\
The goal of this paper is to fill this ambiguity gap and let the fiducial wavefunction be time dependent, extending the semi-classical scheme and ultimately recovering full quantum mechanics. The practical implementation of this construction will be realised by letting the fiducial wavefunction depend on time and application of the variational principle to obtain dynamics. This approach was tested previously for affine coherent states \cite{Malkiewicz:2018}, now the scheme is generalised to the Weyl-Heisenberg coherent states in more cohesive and in-depth way. Moreover the decomposition of the system into `classical' and `quantum' degrees of freedom is introduced, giving an insight of how classical the state is during its evolution. The approach to quantum dynamics and `classical-quantum' split from a variational principle exists in literature \cite{Cesare:2015, Cesare:2016}, but it was not studied from coherent states perspective, except in \cite{Malkiewicz:2018}. The general statements about the interaction between the `classical' and `quantum' degrees of freedom and how the classical trajectories are corrected by quantum effects are made. The motivation for observing the interaction of `classical' and `quantum' degrees of freedom comes from the field of quantum cosmology. Recently, an approximate semi-classical description of dynamics using coherent states (referred to here as the ``lower symbol method'') became popular \cite{Piechocki:2020,Bergeron:2017, klauder2019} and it is vital to understand how accurate it is. The introduced formalism is aimed to answer, in the future, the question of how much quantum effects are able to push the universe from its classical trajectory.\\
Section \ref{sec:1} is a review of a construction of Weyl-Heisenberg coherent states, with a special emphasis on the fiducial vector dependence. Apart from the standard introduction to the topic of generalized coherent states the focus is directed towards the dynamical description of coherent states. The section ends with the example of the dynamical laws governing the evolution of standard Schr\"odinger coherent state.\\
The main part of this paper is section \ref{sec:2} in which the formalism for evolving coherent states is introduced. Starting from an action of quantum mechanics one obtains phase space and hamiltonian in terms of `classical' and `quantum' degrees of freedom. The resulting system is subject to \emph{physical centering} conditions which assure the correct physical interpretation of the dynamics. In the last part of this section the status and interaction between different degrees of freedom is discussed and the definition of a classical state formulated.\\
Section \ref{sec:approximate} focuses on the approximate descriptions of the dynamics in coherent states. The standard, lower symbol, method is reviewed and generalised. An example of a meta-stable state is studied.\\
After that, the paper is concluded.\\ \\
The convention in this paper is set to use natural units in which $\hbar = 1$.

\section{Weyl-Heisenberg coherent states}\label{sec:1}
The first two parts of this section contain a set of selected facts about Weyl-Heisenberg coherent states, for a more comprehensive review of coherent states the reader is encouraged to see, for example \cite{Bergeron:2017II, Gazeau:2009}.
\subsection*{Weyl-Heisenberg group and Lie algebra}
First, define a Weyl-Heisenberg group $WH \simeq \mathbb{R}^3$ with a three-parameter group element $WH \ni g = (s,q,p)$ satisfying group action
\begin{equation}\label{eq:group_action}
(s_1,q_1,p_1) \cdot (s_2,q_2,p_2) = (s_1 + s_2 - \frac{1}{2}(q_1 p_2 - q_2 p_1), q_1 + q_2, p_1 + p_2).
\end{equation}
It is easy to check that a neutral element is $(0,0,0)$ while an inverse element to $g$ reads $g^{-1} = (-s,-q,-p)$.\\
Weyl-Heisenberg Lie algebra $\mathfrak{WH}$ is spanned on three infinitesimal generators $\1, \q, \p$ with basic commutators
\begin{equation}
[\1,\q] = 0, \ [\1,\p] = 0, \ [\q,\p] = i \hbar \ \ (in\ natural\ units\ \hbar = 1).
\end{equation}
A general element of $\mathfrak{WH} \ni X$ can be parametrized as $X = i s \1+ i p \q - p \p$. Using an exponential map on infinitesimal operator $X$ one obtains
\begin{equation}
e^X = e^{is\1+i(p\q-q\p)} = e^{is \1} e^{-\frac{i}{2} qp} e^{ip\q} e^{-iq\p} = e^{is\1} e^{\frac{i}{2} qp } e^{-iq\p} e^{ip\q},
\end{equation}
where the Baker-Campbell-Hausdorff formula was used to obtain the latter expressions.\\
The composition rule reads
\begin{equation}\label{eq:composition_rule}
e^{X_1} e^{X_2} = e^{i(s_1+s_2)\1}e^{-\frac{i}{2}(q_1p_2-p_1q_2)\1}e^{i(p_1+p_2)\q-i(q_1+q_2)\p},
\end{equation} 
which agrees with the Weyl-Heisenberg group action (\ref{eq:group_action}).\\
It is customary to introduce a displacement operator
\begin{equation}
 \D(q,p) \equiv e^X,
\end{equation}
which is unitary and $\D^{-1}(q,p) = \D^{\dagger}(q,p) = \D(-q,-p)$.\\
The displacement operator has a useful property which is used frequently throughout the paper
\begin{equation}\label{eq:displacement_commutation}
\hat{f}(\1,\q,\p) \D(q,p) = \D(q,p)\hat{f}(\1,q\1+\q,p\1+\p).
\end{equation}
As can be seen in formula (\ref{eq:composition_rule}) the $s$ parameter is an overall phase of $D(q,p)$ operator. By itself it has no observational importance (as will be presented in  Sec. \ref{sec:2}) but is needed to satisfy the Schr\"odinger equation (as will be presented in the next part of the current section). Due to the lack of observational importance $s$ is usually suppressed in notation.
\subsection*{Generalized coherent state}
A family of generalised coherent states is understood as states, parametrized by two external parameters $(q, p)$
\begin{equation}
|q,p \rangle \in L^2(\chi),\ (q,p) \in \chi,
\end{equation}
which satisfy the following properties \cite{Gazeau:2009}:
\begin{itemize}
\item \textit{A map $\chi \ni (q,p) \rightarrow |q,p\rangle \in L^2(\chi)$ is continuous}
\item \textit{Coherent states form an overcomplete family of states with the resolution of unity}
\begin{equation}
\1= \int_{\chi}d\mu (q,p) |q,p \rangle \langle q,p |.
\end{equation}
\item \textit{Generalized coherent states are constructed by an action of the unitary irreducible representation of the symmetry group of $\chi$ on some fiducial vector $| \phi \rangle$:}
\begin{equation}
| q,p \rangle = U(q,p) | \phi \rangle,\ \phi \in L^2(\mathbb{R}).
\end{equation}
\end{itemize}
In the case of a Weyl-Heisenberg coherent state, the space of parameters $(q,p)$ is just $\mathbb{R}^2$ with the measure $d\mu(q,p) = \frac{dqdp}{\pi}$. Construction is done with a displacement operator $\D(q,p)$, a unitary irreducible representation of a Weyl-Heisenberg symmetry group
\begin{equation}
| q,p \rangle \equiv | q,p; \phi \rangle \equiv \D(q,p) | \phi \rangle.
\end{equation}
All three expression above are equivalent and are used interchangeably throughout the text. The two latter expressions highlight the dependence of a coherent state on a fiducial vector, which is implicit in the most popular notation $| q,p \rangle$. \\
The position representative of Weyl-Heisenberg coherent states is
\begin{equation}
\langle x | q,p  \rangle =e^{i s} e^{\frac{i}{2} qp} e^{i p x} \phi(x-q).
\end{equation}
One of the most important features of coherent states is the recognition of the physical interpretation of $(q,p)$ parameters.\\
Using relation (\ref{eq:displacement_commutation}) one obtains the following expectation values
\begin{subequations}
\begin{align}
\langle q,p |\q | q,p \rangle &= q + \langle \phi | \q | \phi \rangle \equiv q + \langle \q \rangle,\label{eq:Q_exp_Val}\\
\langle q,p |\p | q,p \rangle &= p + \langle \phi | \p | \phi \rangle \equiv p + \langle \p \rangle,\label{eq:P_exp_Val}
\end{align}
\end{subequations}
where a shorthand notation for an expectation value in a fiducial vector was introduced.\\
As can be seen from the above equations (\ref{eq:Q_exp_Val}-\ref{eq:P_exp_Val}) for the so called \textit{physical centering} conditions \cite{Klauder:2015} $\langle \q \rangle = 0$ and $\langle \p \rangle = 0$ one obtains a clear physical interpretation of the labels $(q,p)$ as a particle's mean position and mean momentum respectively. The above conditions depend directly on the fiducial wavefunctions and in order to have them fulfilled the fiducial space becomes an abstraction class of a Hilbert space $| \phi \rangle \in \Phi = \left( L^2(\mathbb{R}) \right)/\left(\langle \q \rangle \langle \p \rangle \right)$. Having a physical interpretation of the $(q,p)$ variables on a \emph{physical centering} constraint surface, they will be referred to as `classical' degrees of freedom, while any other variables will be called `quantum'. The motivation for this naming is clear, for a fiducial wavefunction being an infinitesimaly narrow wave packet, all `quantum' degrees of freedom vanish. As it will be presented below `classical' variables not only correspond to mean position and momentum but also form a classical phase space.\\

\subsection*{Dynamics and the Schr\"odinger coherent state}
The main subject of this paper is the analysis of the quantum mechanical dynamics of coherent states.
The last part of this section will utilise a very special example of Weyl-Heisenberg coherent state - a Schr\"odinger state - to demonstrate meaning and possible issues of a formalism to be developed.\\
The Schr\"odinger coherent state is constructed by the action of the displacement operator on a harmonic oscillator vacuum $\D(q,p)|0\rangle$,  where
\begin{equation}\label{eq:HO_vacuum}
\langle x | 0 \rangle = \left( \frac{m \omega}{\pi} \right)^{\frac{1}{4}}e^{-\frac{m \omega x^2}{2}}.
\end{equation}
It solves exactly Schr\"odinger equation for a harmonic oscillator hamiltonian $\hat{H}_{HO}(\q,\p) = \frac{1}{2m}\p^2 + \frac{m\omega^2}{2} \q^2$, where $m$ is the particle's mass, $\omega$ is an angular frequency and $\hat{H}_{HO}(\q,\p)|0 \rangle = \omega/2 | 0 \rangle$.\\
Schr\"odinger coherent state has an unique property of stability, meaning that all the quantum dynamics, generated by $\hat{H}_{HO}$ can be incorporated into group action
\begin{equation}\label{eq:stability}
| q,p; 0 \rangle (t) = \D(q(t),p(t))| 0 \rangle.
\end{equation}
Plugging in the state (\ref{eq:stability}) into Schr\"odinger equation one gets
\begin{equation} \label{eq:Schrodinger}
\left[-\dot{s}\1+\dot{q}p\1 - \dot{\left( \frac{qp}{2} \right)}\1 + \dot{q}\p-\q \dot{p} + \1\frac{d}{dt} - \hat{H}_{HO}(q \1 +\q, p \1+\p) \right] | 0 \rangle=0.
\end{equation}
The displaced hamiltonian can be decomposed into the following terms $\hat{H}_{HO}(q\1+\q,p\1+\p) = \hat{H}_{HO}(q\1,p\1)+\hat{H}_{HO}(\q,\p)+p\p/m+\omega^2 m q \q$ leading to the convenient arrangement of the terms in the above equation
\begin{equation}\label{eq:HO_schrodinger_hamiltonian}
\left[ \dot{q}p - \dot{\left( \frac{qp}{2}\right)} - \hat{H}(q,p) \right] |0 \rangle + \left[-\dot{s}-\frac{\omega}{2}  \right] | 0 \rangle+\left[ -\dot{p}-m\omega^2 q \right]\q | 0 \rangle+\left[ \dot{q}-\frac{p}{m} \right]\p | 0 \rangle=0.
\end{equation}
Observe that the fiducial vector (\ref{eq:HO_vacuum}) satisfies the \emph{physical centering} conditions $\langle 0 | \q |0 \rangle=0$, $\langle 0 | \p |0 \rangle=0$, therefore both vectors $\q | 0 \rangle$ and $\p | 0 \rangle$ are orthogonal to $| 0 \rangle$. It means  that the last two terms of equation (\ref{eq:HO_schrodinger_hamiltonian}) have to vanish independently from others.\\ 
As $[\q,\p] \neq 0$, vectors $\q | 0 \rangle$ and $\p | 0 \rangle$ in general cannot be made mutually orthogonal, but in the special case of gaussian fiducial vector (\ref{eq:HO_vacuum}) they are parallel to each other $\p| 0 \rangle = i m\omega \q | 0 \rangle$, leading to the classical hamilton equations for parameters $(q,p)$
\begin{subequations}
\begin{align}
\dot{q} &= \frac{\partial H_{HO}(q,p)}{\partial p} \label{eq:hamilton_q},\\
\dot{p} &= - \frac{\partial H_{HO}(q,p)}{\partial q} \label{eq:hamilton_p},
\end{align}
\end{subequations}
where the two equations were obtained due to \emph{complex polarization} \cite{Klauder:2015} of the terms proportional to $\q | 0 \rangle$, $\p | 0 \rangle$.
Plugging in the above equations (\ref{eq:hamilton_q}-\ref{eq:hamilton_p}) to (\ref{eq:HO_schrodinger_hamiltonian}) and setting $s=-\omega/2\ t$ solves the Schr\"odinger equations.\\ \\
From the perspective of possibility of generalizing the description of dynamics in coherent states to any hamiltonians and fiducial vectors, the following conclusions from the above example prove to be useful:\\
- By setting \emph{physical centering} conditions, one obtains equations for $(q,p)$ which resemble hamilton equations.\\ 
- If the fiducial vectors were to evolve, to maintain the physical interpretation of $(q,p)$, one would have to impose \emph{physical centering} constraints on fiducial vector at all times.\\
- The classical equations for $(q,p)$ were obtained from mixed terms in the decomposition of the hamiltonian $\hat{H}_{HO}$. For different hamiltonians, mixed terms could lead to modified, hamilton-like equations.\\
- The expressions proportional to fiducial vector $\propto | 0 \rangle$ (first two terms in (\ref{eq:HO_schrodinger_hamiltonian})) resemble a Legendre transform and one might expect that parts of them will be a seed for a `classical-quantum' phase space.\\
From this point on, the use of identity operator $\1$ in the formulae will be suppressed, as its presence is obvious from the context, but obscures the clarity of notation.
\section{The decomposition in a canonical formalism}\label{sec:2}
Having introduced Weyl-Heisenberg coherent states and analysed the laws of motion for Schr\"odinger coherent state the general formalism for quantum dynamics in coherent states is proposed in this section. The approach starts from the action for quantum mechanics restricted to coherent states and follow with the canonical treatment of constrained systems.
\subsection*{The action and variational principle}
The quantum mechanical action
\begin{equation}\label{eq:QM_action}
S = \int dt \langle \psi | i \frac{d}{dt} - \hat{H}(\q,\p) | \psi \rangle,\ \ |\psi \rangle \in L^2(\mathbb{R},dx),
\end{equation} 
is normally varied with respect to the amplitudes $\psi(x) \in \mathbb{C}$ at $x \in \mathbb{R}$, which admits no classical interpretation. The idea is to use the Weyl-Heisenberg coherent states $| q,p; \phi\rangle$ instead of $|\psi \rangle$ and vary the quantum action with respect to the parameters $(q, p)$ and the fiducial amplitude $| \phi \rangle$ independently. The variables $q$ and $p$ would admit the classical interpretation as `position' and `momentum', where the fiducial vector $| \phi \rangle$ would be responsible for higher order contribution to the dynamics related to the shape of the underlying wavefunction. This idea was employed first for affine coherent states in \cite{Malkiewicz:2018} and will be followed in this paper for Weyl-Heisenberg states.\\
Using the results from previous section one can immediately obtain the quantum action for a Weyl-Heisenberg coherent state,
\begin{equation}\label{eq:cs_action}
S=\int dt \left[-\dot{s}+ \dot{q}p - \dot{\left( \frac{qp}{2} \right)}+ \langle \phi |i \frac{d}{dt} | \phi \rangle +\dot{q} \langle \phi |\p| \phi \rangle -\dot{p} \langle \phi |\q| \phi \rangle - \langle \phi | \hat{H}(q+\q,p+\p) | \phi \rangle \right] ,
\end{equation}
where the terms $\dot{\left( \frac{qp}{2} \right)}$ and $\dot{s}$ can be disregarded as they correspond to vanishing boundary terms.\\
The phase space of the above action is described in terms of the following symplectic form
\begin{equation}
\omega = dq \wedge d(p + \langle \q \rangle ) + d \langle \q \rangle \wedge dp + \int dx\ d \phi(x) \wedge d(i \phi^*(x)),
\end{equation}
where the expression with explicit dependence on $\phi(x)$ can be recognized as defining the standard canonicaly conjugate pair of a ``Schr\"odinger field''. A practical realisation of the phase space of fiducial vectors on finite dimensional Hilbert space is presented in Appendix \ref{Ap:A}. From now on the term $\int dx\ d \phi(x) \wedge d(i \phi^*(x))$ will be denoted as $i\ d| \phi \rangle \wedge d  \langle \phi |$.\\
At this point the kinetic part of the action (\ref{eq:cs_action}) contains terms which mix `classical' degrees of freedom $(q,p)$ and `quantum' terms which depend on the fiducial vector, therefore it is natural to make the transformation $q \mapsto q- \langle \q \rangle,\ p \mapsto p- \langle \p \rangle$ leading (up to boundary terms) to 
\begin{subequations}
\begin{align}
S &= \int dt \left[ \dot{q}p - \langle \dot{\q} \rangle \langle \p \rangle + \langle i \frac{d}{dt}  \rangle - \langle \hat{H} \left(q-\langle \q \rangle+\q,p-\langle \p \rangle+\p \right) \rangle  \right],  \\
\omega &= dq \wedge dp - d\langle \q \rangle \wedge d \langle \p \rangle + i\ d| \phi \rangle \wedge d \langle \phi |,
\end{align}
\end{subequations}
and effectively decoupling the `classical' variables $(q,p)$ on the kinetic space from other degrees of freedom.
Observe that, on a constraint surface set by the physical centering conditions $\langle \q \rangle =0$ and $\langle \p \rangle =0 $, the `classical' variables $(q,p)$ coincide before and after transformation. Although at first sight one is tempted to claim that the transformation separated the constraints from other degrees of freedom, the part $i\ d| \phi \rangle \wedge d \langle \phi |$ contains implicitly $\langle \q \rangle$ and $\langle \p \rangle$. Therefore one is forced to apply a constrained system's theory for a further dynamical analysis.\\
\subsection*{Canonical analysis of constrained system}
Choosing to follow a canonical analysis one obtains a total hamiltonian
\begin{equation}\label{eq:tot_hamiltonian}
\langle \hat{H}_T \rangle = \langle \hat{H} \left(q-\langle \q \rangle+\q,p-\langle \p \rangle+\p \right) \rangle + \alpha \langle \q \rangle + \beta \langle \p \rangle,
\end{equation}
where $\alpha$ and $\beta$ are Lagrange multipliers. The dynamics is driven by the following compound bracket
\begin{equation}\label{eq:compound_bracket}
\frac{d}{dt} \langle \phi | \hat{O}(q,p) | \phi \rangle = \llbracket \langle \hat{O}(q,p) \rangle , \hat{H}_T \rrbracket \approx \{\langle \hat{O}(q,p) \rangle,\langle \hat{H}_T \rangle \}_{qp}-i \langle [\hat{O}(q,p),\hat{H}_T] \rangle,
\end{equation}
where the weak equality sign ``$\approx$''  is translated as ``equal on the constraint surface''. The first term contains a standard Poisson Brackets $\{f,g \}_{xy}=(\partial f/\partial x) (\partial g/\partial y) - (\partial f/\partial y) (\partial g/\partial x)$ and the last term contains a commutator.\\
As the operators $\q$ and $\p$ do not commute, the constraints are clearly of a second class. One should now check the consistency relations for the constraints and continue the analysis with either introducing a Dirac bracket instead of commutator or trying to solve the Lagrange multipliers. The latter of those two equivalent choices will be presented in this paper.\\
Checking consistency conditions for constraints one obtains
\begin{subequations}
\begin{align}
\frac{d}{dt} \langle \q \rangle &\approx 0 \Rightarrow \beta \approx - \left\langle \frac{\partial \hat{H}\left(q+\q,u \right)}{\partial u}\bigg\rvert_{u=p+\p} \right\rangle, \label{eq:beta}\\
\frac{d}{dt} \langle \p \rangle &\approx 0 \Rightarrow \alpha \approx - \left\langle \frac{\partial \hat{H}\left(u,p+\p \right)}{\partial u}\bigg\rvert_{u=q+\q} \right\rangle. \label{eq:alpha}
\end{align}
\end{subequations}
Having obtained and fixed expressions for Lagrange multipliers $\alpha, \beta$ one is allowed to generate the motion of the system unambiguously, using the total hamiltonian $\langle \hat{H}_T \rangle$ and the compound bracket (\ref{eq:compound_bracket}). The constraints are kept constant during evolution and there is no need of introducing any further conditions.\\ \\
The first expression in the equation (\ref{eq:tot_hamiltonian}) is a physical hamiltonian. After solving the constraints one sees that from the point of view of dynamics on the constraint surface the constraints can be omitted in the arguments of the physical hamiltonian 
\begin{equation}
\langle \hat{H}(q-\langle \q \rangle+\q,p-\langle \p \rangle+\p) \mapsto \hat{H}(q+\q,p+\p) \rangle.
\end{equation}
Assuming that the physical hamiltonian is polynomial in the first and quadratic in the second argument there exists a convenient decomposition
\begin{equation}\label{eq:H_decomposition}
\langle \hat{H}(q+\q,p+\p) \rangle = \underbrace{\frac{1}{2m} p^2 + V(q)}_{H_C} + \underbrace{\langle \frac{1}{2m} \p^2 + V(\q) \rangle}_{\langle \hat{H}_Q \rangle} + \underbrace{\frac{p}{m} \langle \p \rangle + \langle V_I(q,\q) \rangle}_{\langle H_I \rangle},
\end{equation}
where the potential
$$ V(q+\q) = \sum_{n=0}^{\infty} c_n (q+\q)^n,$$
and
$$V_I(q,\q) = \sum_{n=0}^{\infty} c_n \sum_{m=1}^{n-1} {n \choose m} q^m \q^{n-m}.$$
It is worth mentioning that in the literature \cite{Gazeau:2009} the expectation values of operators in coherent states are referred to as lower symbol operators $\check{O} = \langle q,p | \hat{O} | q,p \rangle$. Above, the physical hamiltonian $\langle q,p |\hat{H}(\q,\p)| q,p \rangle = \langle \phi | \hat{H}(q+\q,p+\p) | \phi \rangle = \check{H}(q,p,\q,\p)$ is a lower symbol hamiltonian.
Throughout the paper $H_C$ will be related as a classical hamiltonian, $\langle \hat{H}_Q \rangle$ as a quantum hamiltonian and $\langle H_I \rangle$ as a interaction hamiltonian with an interaction potential $ \langle V_I \rangle$. Observe that only constraints and interaction hamiltonian mix `classical' and `quantum' degrees of freedom.\\
Using the newly introduced notation one can write equations of motion in a convenient way,
\begin{subequations}
\begin{align}
\dot{q} &\approx \frac{\partial H_C(q,p)}{\partial p} \approx \frac{p}{m} \label{eq:HL-q}\\
\dot{p} &\approx -\frac{\partial V(q)}{\partial q} - \frac{\partial \langle V_I(q,\q) \rangle}{\partial q} \label{eq:HL-p}\\
\frac{d}{dt}\langle f(\q,\p) \rangle &\approx -i \langle [f(\q,\p),\hat{H}_Q] \rangle -i \langle [f(\q,\p),\hat{H}_I + \alpha \q + \beta \p] \rangle \label{eq:HL-Quant}
\end{align}
\end{subequations}
The equations (\ref{eq:HL-q}) and (\ref{eq:HL-p}) describe evolution of `classical' variables $(q,p)$. As they are of the form similar to hamilton equations (with a difference of the additional interaction term), therefore they will be referred to as hamilton-like equations. The interaction term in the hamilton-like equations pushes the evolution of $(q(t),p(t))$ from its classical trajectory, correcting it by accounting for quantum effects. Equation (\ref{eq:HL-Quant}) describes evolution of any function of basic operators $\q$ and $\p$. Again, it is worth noticing that motion is not only generated by a quantum hamiltonian $H_Q$ but also by the interaction term and constraints.\\
The presented formalism can be viewed as treating the mean position and momentum on a special footing, therefore extracting it from standard quantum mechanical dynamics. This effectively leads to describing the evolution of the shape of a fiducial wavefunction from a semi-classical frame following the `classical' degrees of freedom $(q,p)$. The formalism is fully equivalent to standard quantum mechanics as long as the fiducial space $\Phi$ densely covers the abstraction class of a quotient space created by dividing out the mean position and momentum from a Hilbert space $\left( L^2(\mathbb{R},dx) \right) /\left( \langle \q \rangle \langle \p \rangle \right)$ of a studied problem.\\ \\
It is instructive now to go back to the example of a Schr\"odinger coherent state. Comparing the hamilton-like equations (\ref{eq:HL-q}) and (\ref{eq:HL-p}) with the equations of motion (\ref{eq:hamilton_q}) and (\ref{eq:hamilton_p}) one can easily recognize the reason why the `classical' and `quantum' degrees of freedom decoupled. For a harmonic oscillator the only two interaction terms $\frac{p}{m} \langle \p \rangle$ and $ m \omega^2 q \langle \q \rangle$ are proportional to constraints and therefore vanish on the \emph{physical centering} surface. In fact, it is easy to see that for systems with free or quadratic potential hamilton-like equations (\ref{eq:HL-q} - \ref{eq:HL-p}) reduce exactly to classical hamilton equations, no matter what fiducial vector is used. For any potential of order three or more, there will be always a correction to the classical motion.\\
Regarding the equation (\ref{eq:HL-Quant}) for a Schr\"odinger coherent state, the first term trivializes due to the fiducial vector $| \phi \rangle = | 0 \rangle$ being an energy eigenstate, while the interaction terms cancel constraints, leading to $d/dt \langle f(\q,\p) \rangle = 0$.\\
\subsection*{Classical states}
Having defined the decomposition (\ref{eq:H_decomposition}) and obtained equations of motion (\ref{eq:HL-q}-\ref{eq:HL-Quant}) it is straightforward to now give an interpretation to $H_C,\ \langle H_Q \rangle,\ \langle H_I \rangle$ terms. The first term, $H_C$, is a classical hamiltonian of a given system, it involves only the classical variables q and p. The second term, i.e. the expectation value of a quantum hamiltonian in the fiducial vector $\langle H_Q \rangle$, is hamiltonian for `quantum' degrees of freedom which are hidden from the classical point of view. The third term, $\langle H_I \rangle$, describes a coupling between classical and quantum degrees of freedom and its vanishing is certainly one of the requirements for the fundamentally quantum system to be in a classical state. A part of the total hamiltonian (\ref{eq:tot_hamiltonian}) are constraints which define a \emph{physical centering} surface. The Lagrange multipliers $\alpha$ and $\beta$, given in (\ref{eq:beta},\ref{eq:alpha}), are in general both $(q,p)$ and $\q, \p$ dependent and thus, constraints also contribute to the `quantum-classical' interaction. \\ \\
For the sake of the discussion below it is convenient to assume that the potential of the system is bounded from below and its minimum value, for some $q_0$, is zero, $V_{min} = V(q_0) = 0$. Therefore one has $H_C \geq 0$, $H_Q >0$ at all times. In general, $V_I$ can be negative.\\ \\
Given above, one is in a position to propose a criterion for the quantum system to be in a classical state \cite{Malkiewicz:UnPub}. First, note the following:
\begin{enumerate}
\item The ``quantum energy'' stored in the `quantum' degrees of freedom should be negligible compared to the ``classical energy'' stored in the `classical' degrees of freedom, i.e. 
\begin{equation}
\langle H_Q \rangle \ll H_C
\end{equation}
\item The interaction potential should be also negligible compared to the ``classical energy'' stored in the `classical' degrees of freedom, i.e.
\begin{equation}
| \langle V_I \rangle | \ll H_C
\end{equation}
\item The constraint term sources a force which acts normally to the motion of the particle and does not contribute to the energy of the system in any way. The constraints are, moreover, on the `quantum' degrees of freedom and they exert a force on the quantum degrees of freedom, only despite that $\alpha$ and $\beta$ may
depend on $q$ and $p$. Therefore, this terms does not affect the classical dynamics of the system.
\end{enumerate}
The above remarks combine together into the following criterium for classicality:\\ \\
\textit{A quantum system is in a classical state when the quantum energy stored in the shape of its wavefunction and its quantum-classical interaction potential are negligible compared to the classical energy stored in the classical degrees of freedom,}
\begin{equation}\label{eq:classicality}
|\langle H_Q \rangle + \langle V_I \rangle| \ll H_C
\end{equation}
Given a specific potential term, one may be able to determine the phase space domains of strongly classical and strongly quantum states. The above definition immediately sparks interesting questions, such as whether a system can move between these domains back and forth, or whether asymptotic behaviour may remain within a fixed domain, let it be a classical or quantum one.\\ \\
Inspired by the above definition one can introduce some measure of `how classical' a state at some given time is. In the case when one is interested mostly in how quantum effects push the `classical' variables from their classical trajectory (see eq. (\ref{eq:HL-p})) one can introduce the ``interaction index''
\begin{equation}\label{eq:interaction_index}
I_I = \bigg\rvert \frac{\langle V_I \rangle}{H_C + \langle \hat{H}_Q + V_I \rangle} \bigg\rvert.
\end{equation}
The above index will be discussed in the example of meta-stable state in Section \ref{sec:approximate}.\\ \\
At this point the investigation of exact quantum mechanical problems is abandoned. The rest of the paper focuses on approximate methods. Standard approaches for investigating semiclassical dynamics using coherent states will be introduced and will be related to the above formalism. As will be presented the standard methods will fit into the formalism as a special case and will be easily extended.\\

\section{Approximate dynamics}\label{sec:approximate}
Usually the methods of approximate description of quantum dynamics rely on the restriction of the available Hilbert space. The dynamical laws are derived from the stationary point of the action (\ref{eq:QM_action}) obtained by a variation with respect to the available degrees of freedom
\begin{equation}\label{eq:QM_action_variation}
\delta \left( \int dt \langle \psi | i \partial_t - \hat{H} | \psi \rangle \right) = 0 \Rightarrow i \partial_t | \psi \rangle = \hat{H} | \psi \rangle.
\end{equation}
Suppose that the quantum action becomes confined to the states living in the subspace $|\psi_{\Gamma} \rangle \in \Gamma \subset \mathcal{H}$. For such reduced
action the wavefunctions $| \psi \rangle_{\Gamma}$ obey the following equation
\begin{equation}\label{eq:Schrodinger_approx}
\langle \delta \psi_{\Gamma} | i \partial_t - \hat{H} | \psi_{\Gamma} \rangle = 0
\end{equation}
which can be translated into the statement that the Schr\"odinger equation holds only pointwise in the tangent
space to $| \psi_{\Gamma} \rangle$
\begin{equation}
T_t\Gamma = span \left( \frac{\partial | \psi_{\Gamma} \rangle}{\partial t}\bigg\rvert_{t} \right)
\end{equation}
In general $T_t\Gamma \neq \Gamma$. As the size of $\Gamma$ is increased the approximation (\ref{eq:Schrodinger_approx}) to Schr\"odinger equation (\ref{eq:QM_action_variation}) becomes better and in the case of $\Gamma = \mathcal{H}$ by the virtue of Parseval’s identity it converges at each point $t$ to full Schr\"odinger equation. As the interest of this paper is the time evolution of the Weyl-Heisenberg coherent states, the general states $| \psi \rangle$ are restricted to $|q,p;\phi \rangle$. In that case the states on the fiducial space are to be effectively restricted to $\Gamma_{\Phi} \subset \Phi$.\\ \\
\begin{figure}[h]
\centering
\includegraphics[width=0.6\textwidth]{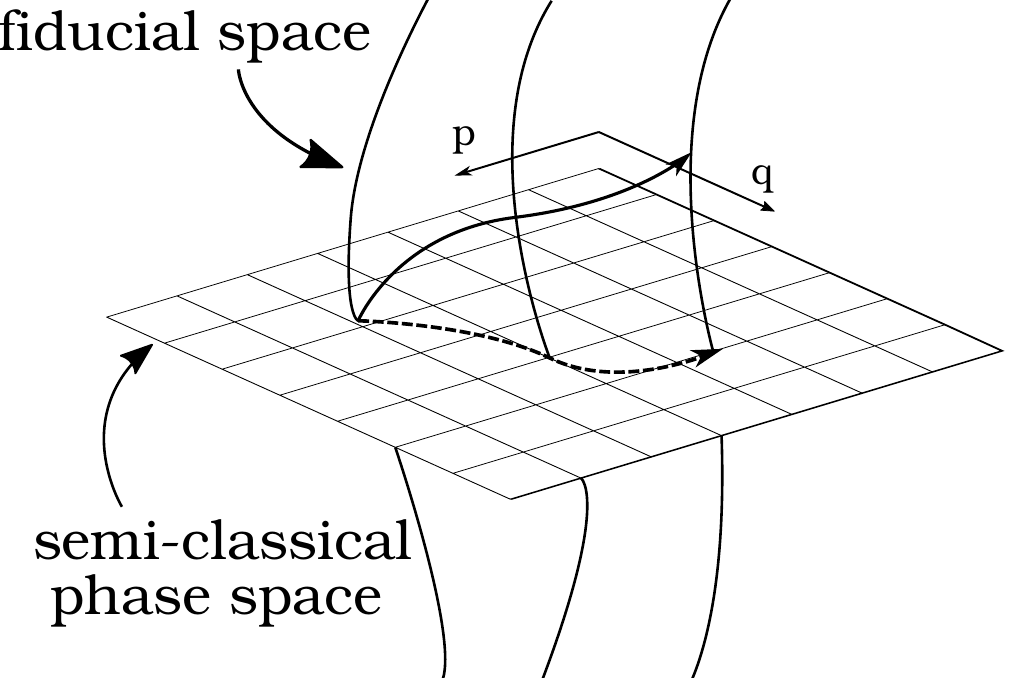}
\caption{A schematic depiction of a dynamical trajectory of a quantum system in a coherent state. The horizontal grid represents a semiclassical phase space for a fixed fiducial vector. A trajectory plotted with a solid line represents unitary evolution driven by a full quantum mechanical action (\ref{eq:cs_action}), while the dashed line trajectory represents motion obtained by the lower symbol approach. Although both motions begin in the same point of semiclassical phase space and fiducial space, they quickly diverge from one another.}
\label{fig:space_foliation}
\end{figure}

\subsection*{Lower symbol method}
In literature the most common approach to obtain semiclassical dynamics using coherent states is to take a fiducial space consisting only of one element $\Gamma_{\Phi} = \{| \phi \rangle\}$ \cite{Piechocki:2020,Bergeron:2017, klauder2019}. Effectively, one calculates the lower symbol hamiltonian $\langle q,p | \hat{H} |q,p \rangle = \check{H}$ at some initial time and generates motion using hamilton-like equations
\begin{subequations}
\begin{align}
\dot{q} &= \frac{\partial \check{H}}{\partial p}, \\
\dot{p} &= -\frac{\partial \check{H}}{\partial q},
\end{align}
\end{subequations}
while keeping all expectation values of functions of $\q$ and $\p$ constant.
Such choice has some appealing properties, mostly connected to simplicity of such description. As the fiducial space is one-dimensional there is no motion in `quantum' degrees of freedom and the shape of the wavefunction is \textit{frozen} in time. That leads to huge simplification, if the \emph{physical conditions} are satisfied on initial data surface, then they cannot leave it during the evolution. The remaining degrees of freedom are $(q,p)$, and therefore all motion takes place on a \textit{semiclassical phase space}. Observe that the hamilton-like equations do not, in general, reduce to classical hamilton equations, as the interaction potential $\langle V_I \rangle$ does not vanish. The interaction potential $\langle V_I \rangle$ is constant in `quantum' degrees of freedom but still corrects the classical motion. What effectively happens is that a given classical hamiltonian is exchanged to different (semi)classical hamiltonian which is believed to describe the dynamics of $(q,p)$ in a better fitted way for a starting fiducial vector.\\
On the other hand, certain aspects of this approach arouse suspicion. The stiff transport of the constant fiducial vector (as can be seen in the Fig. \ref{fig:space_foliation}) via the action of the group representation $D(q,p)$ is clearly far from what really happens in the full system. One has to expect that, except very particular systems, the trajectory computed by this method quickly diverges from the true motion. Worse, there is no natural control parameter which could estimate how good the approximation is during the evolution.

\subsection*{Basis truncation method}
The standard extension of the lower symbol method is carried out by directly increasing the basis of the fiducial space and adding more orthonormal vectors to it, $\Gamma_{\Phi} = \{ | \phi_1 \rangle, | \phi_2 \rangle, \dots , | \phi_N \rangle \}$. The practical realisation of a computational scheme for this method is presented in the Appendix \ref{Ap:A}. Now the `quantum' degrees of freedom can evolve within $\Gamma_{\Phi}$ and comparing with with the previous approach one obtains corrected motion, much more similar to the solid line trajectory in Fig. \ref{fig:space_foliation}. With smart choice of basis, this method can give a really good, but computationally demanding, approximation of quantum motion. For a good choice of fiducial space basis it is easy to control the accuracy of the method by monitoring a share of a particular state in the resulting fiducial vector.
\subsection*{Moments expansion method}
Another possible extension of the lower symbol method is the moments expansion. The method is well known for years and becomes reinvented every once in a while in context of different fields \cite{Prezhdo:2000, Pahl:2002, Bayta:2019, Bayta:2020}. It introduces the notion of moments of the wave function. If the moment, for example $\langle \psi | \q^n \p^m | \psi \rangle$, is not centered, then can be written as a sum of moments up to order $(n+m)$. The essential assumption is that the moments satisfy a hierarchy of orders, where moments of higher orders contribute less to the dynamics. If so, the equations of motion can be expanded up to some order in moments and then truncated by imposing that all higher order moments vanish. By going to higher and higher orders one implicitly expands the space $\Gamma_{\Phi}$.\\
Observe that the introduced formalism for dynamics in Weyl-Heisenberg coherent states is well suited for this kind of approximation. The physical centering conditions impose that the `quantum' degrees of freedom $\langle f(\q,\p) \rangle$ are already central moments and can be readily used for expanding the equations of motion. The `classical' variables $(q,p)$ are of first order.\\
The approximation scheme is clearly described in the QHD algorithm \cite{Prezhdo:2000} (adapted to the formalism presented in this paper):
\begin{enumerate}
\item \textit{Variables} - Define the set of all degrees of freedom, including `classical' variables and `quantum mechanical' observables up to some order.
\item \textit{Total hamiltonian} - Find the total hamiltonian (\ref{eq:tot_hamiltonian}) in terms of classical and quantum degrees of freedom.
\item \textit{Equation of motion} - Generate the equations of motion for the `classical' variables and the fiducial expectation values for the observables from the set defined in Step 1.
\begin{itemize}
\item If the equations of motion do not generate higher order moments, solve the equations
\item If the equations of motion generate higher order moments apply the closure scheme and solve the equations. In the presented formulation the closure scheme is conveniently defined, one just puts higher order moments to zero.
\end{itemize}
\end{enumerate}
The accuracy of this method is estimated by checking if the hierarchy of orders is maintained. Another good control parameter is the constancy of the total hamiltonian.

\subsection*{Example: Meta-stable state}
%
\begin{figure*}[ht]
\centering
\includegraphics[width=\textwidth]{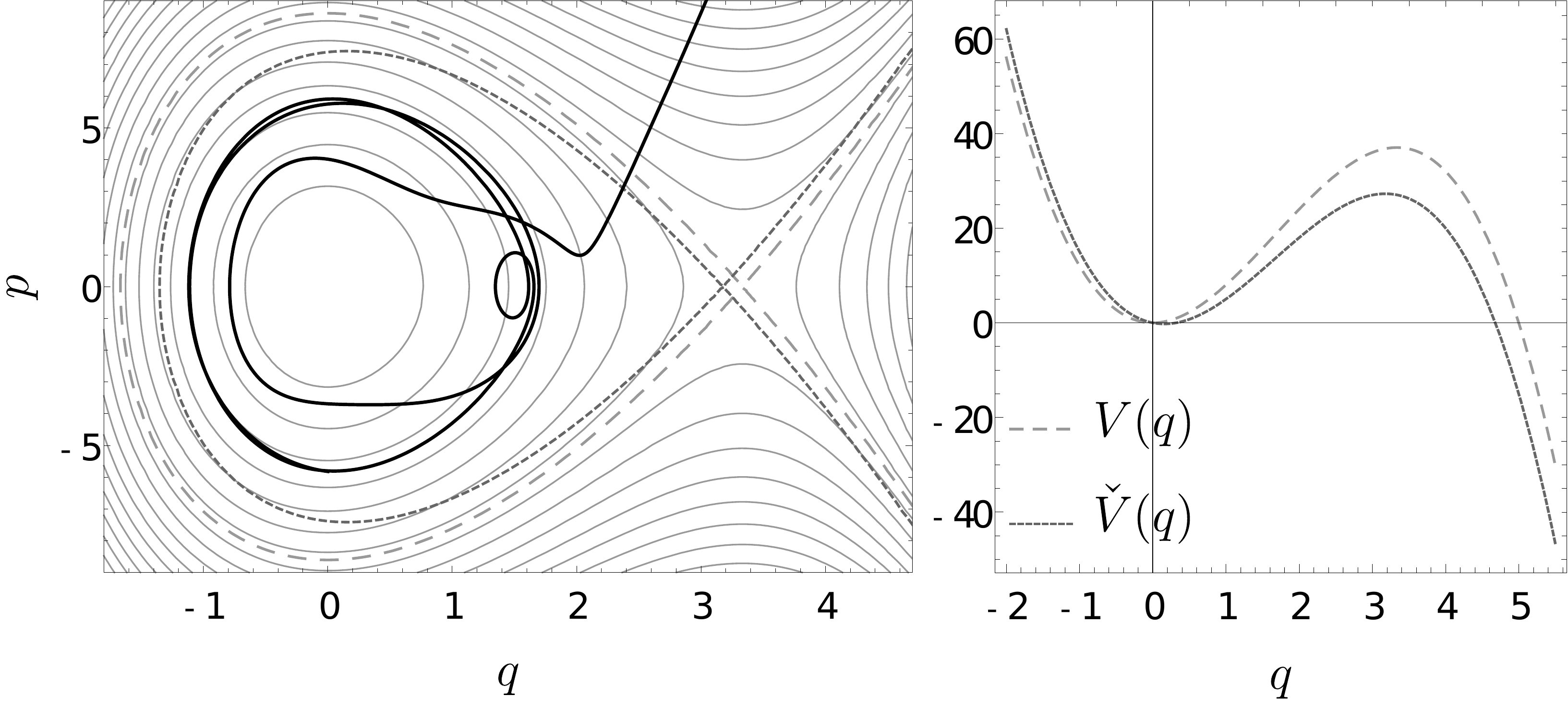}
\caption{The simulations for $a_2=10$, $a_3 = -2$ and the fiducial state $| \phi \rangle = | 0 \rangle$ (\ref{eq:HO_vacuum}) with $m\omega=1$.\\
\emph{Right:} The classical potential $V(q)$ (long-dashed line) and effective potential $\check{V}(q)$ (short-dashed line). The energetic barrier for the effective potential is lowered, therefore in lower symbol method approximation some initial values for classically trapped states lead to the state escaping through the potential barrier.\\
\textit{Left:} Phase space portrait (solid black line) for a particle with initial position $q_0 = 0$ and initial momentum $p_0=-5.82$.
The gray solid lines are the isoenergetic contours for a classical problem, with the energy of the barrier marked by long-dashed contour. The short-dashed line indicates the energy of the barrier in the lower symbol method.
}
\label{fig:example1}
\end{figure*}
A typical example problem for an approximate semi-classical analysis is a system of a particle which, classically, is trapped in the potential's local minimum but is able to escape the meta-stable state by the quantum-mechanical tunnelling effect. The studied model includes a potential of cubic order
\begin{equation}
\label{example_system}
\begin{split}
\hat{H} &= \frac{1}{2} \hat{P}^2 + V(\hat{Q}),\\
V(x) &=a_0 + a_1 x + a_2 x^2+a_3 x^3
\end{split}
\end{equation} 
For simplicity, the parameters $a_0$ and $a_1$ will be chosen to vanish while $a_2 >0$ and $a_3<0$.\\
The classical particle is trapped in the potential's local minimum around $x=0$, where its energy is between $0>E> \frac{4}{27} \frac{a_2^3}{a_3^2}$. The particle exceeding this energy can cross the potential barrier and ultimately escape to infinity. Quantum mechanically, the particle is trapped for a finite amount of time, and one expects that even if the wavefunction of the particle is well localised inside the potential's local minimum it will tunnel outside the trapping potential eventually.\\
The decomposed total hamiltonian (\ref{eq:tot_hamiltonian}) with Lagrange multipliers solved reads
\begin{equation}
H =\underbrace{ \frac{1}{2}p^2 + a_2 q^2 + a_3 q^3 }_{H_C}  + \underbrace{ \langle \frac{1}{2}\p^2 + a_2 \q^2 + a_3 \q^3\rangle }_{\langle \hat{H}_Q \rangle} + \underbrace{ 3 a_3 q \langle \q^2 \rangle -3 a_3 \langle \q \rangle \langle \q^2 \rangle}_{\langle V_I \rangle + \alpha \langle \q \rangle + \beta \langle \p \rangle}
\end{equation}
The hamilton-like equation (\ref{eq:HL-q}-\ref{eq:HL-p}) read
\begin{equation}
\begin{split}
\dot{q}&\approx p\\
\dot{p}&\approx -\left(3 a_3 \langle \q^2 \rangle +2 a_2 q +3 a_3 q^2 \right)
\end{split}
\end{equation}
One can easily see that there is a quantum correction in the equations of motion related to the spread of the fiducial wavefunction. From the point of view of `classical' degrees of freedom, they evolve in the effective potential 
\begin{equation}
\label{eq:effective_potential}
\check{V}(q)= 3 a_3 \langle \q^2 \rangle + a_2 q^2 + a_3 q^3
\end{equation}
One can immediately see that as $\langle \q^2  \rangle$ is always positive the correction is always lowering the potential barrier so the particle which classically is prohibited from escaping, semi-classically can escape the trapping potential. This feature can be seen schematically in the right part of FIG. \ref{fig:example1}. Observe that the lowering of the potential depends on the spread of the wavefunction of the particle. The particles well localised in the potential (with small spread comparing to the width of the trapping potential) are likely to stay trapped but the particles with high spread will \textit{feel} the space outside the potential and escape. This behaviour agrees with the standard intuition.\\ \\
The left side of FIG. \ref{fig:example1} presents a phase space portrait for a particle trapped in the local minimum at initial time. The initial values are chosen such that both in the case of classical system and lower symbol method the particle would be trapped eternally. One can see that after couple of revolutions inside the trapping potential the particle escapes outside. The evolution is obtained by using the moments expansion method. One should not be worried by the crossing of the particle's trajectory with itself. The motion takes place both in `classical' and `quantum' variables while the portrait show only projection of the trajectory to $(q,p)$ variables. In full phase space the trajectories do not cross.\\
\begin{figure*}[ht]
\centering
\includegraphics[width=0.7\textwidth]{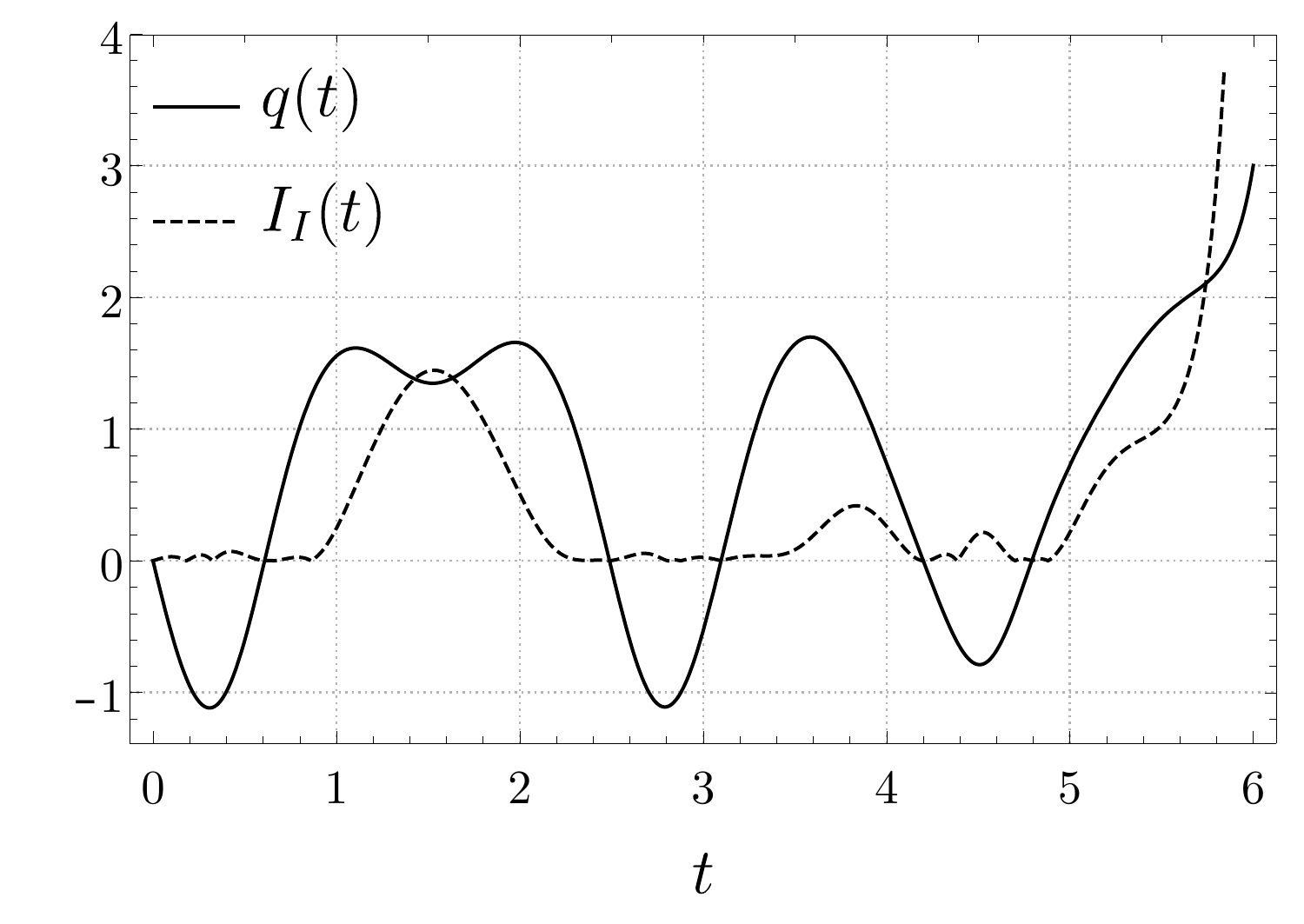}
\caption{Comparison between the evolution of `classical' position $q(t)$ and the interaction index $I_I$ (\ref{eq:interaction_index}) for the same simulation as on FIG. \ref{fig:example1} ($a_2=10$, $a_3 = -2$, $| \phi \rangle = | 0 \rangle$, $q_0 = 0$, $p_0=-5.82$). The noticeable increases in $I_I(t)$ correspond to points in evolution where the semi-classical trajectory diverges from classical one.}
\label{fig:interaction_index}
\end{figure*}
FIG. \ref{fig:interaction_index} shows the relationship between `classical' position $q(t)$ and the interaction index introduced in (\ref{eq:interaction_index}). For initial values taken for this simulation one expects that motion of classically trapped particle would be represented by a periodic function in its position. One can see at least two moments in evolution when the interaction index increases rapidly. First time it happens around $t=1.5$ when a particle almost tunnels through a barrier, but eventually returns to the oscillating trajectory. Second time a noticeable increase in the interaction index can be seen after $t=5$, and this time the particle actually escapes the potential barrier, the motion diverges vastly from the motion of a classical particle. As was anticipated the interaction index $I_I(t)$ highlights the periods of evolution of quantum system when the motion is pushed away from the classical trajectory due to quantum effects.

\section{Conclusions}
The work presented in this paper starts with the definition of generalized coherent states and develops full description of quantum mechanical evolution in Weyl-Heisenberg coherent states. In the introduced formalism the dynamics is presented from a semi-classical frame, attached to the quantum-corrected trajectory of the `classical' degrees of freedom $(q(t),p(t))$. Due to the \emph{physical centering} conditions the labels $(q,p)$ are interpreted as a classical position and momentum and form a canonically conjugate pair over a symplectic manifold. The `quantum' degrees of freedom are connected to the fiducial vector which, in the introduced formalism, also evolves. They are strictly responsible for the shape of the wavefunction transported during evolution with a Weyl-Heisenberg group action $\D(q(t),p(t))$.\\
The novelty of the introduced formalism is that now, all coherent states are able to solve the Schr\"odinger equation exactly. It comes with a price of introducing two second-class constraints to the system. They effectively reduce the dimension of the fiducial space to ascertain the meaning of additionally introduced variables $(q,p)$.\\
With a system consisting of `classical' and `quantum' degrees of freedom it is natural to divide it into classical, quantum and interaction parts. The decomposition leads to the definition of a classical state as one in which the classical part dominates. The classical and quantum parts of the system could evolve separately, without any influence on each other, but the interaction part constantly mixes them and the impact of the `quantum' degrees of freedom on `classical' ones is evident, and \emph{vice versa}. In any system in which the potential, expanded in polynomials, is of order higher than two, the interaction part is present. This is the origin of the correction of the classical motion by `quantum' degrees of freedom.
The introduction of `classical' and `quantum' degrees of freedom led to new definition of a classical state. An interaction index, a measure of how strong is the influence of quantum effects on classical trajectory was proposed.\\
Additionally, the formalism was related to the standard coherent state semi-classical methods of approximate analysis of quantum systems. The standard lower symbol method was recognized as a special case of the full formalism and extensions of it were proposed. Contrary to the standard method, the extensions have a natural possibility to control the accuracy of the simulated motion. The extended schemes have a low accuracy limit coming down to the lower symbol method and the high accuracy limit of, effectively, full quantum mechanics.\\
Although the paper closes one mathematical gap in the toolkit of coherent states applied to physics, it also generates a multitute of questions regarding quantum systems. Are there other families of coherent states that fit the semiclassical analysis of certain systems better? Is the flow between phase space domains of classical and quantum states described by some additional laws? Can a modified trajectory return to its classical analogue? What are the asymptotics of the dynamics of quantum and classical states?\\
Those questions seem to be especially important from the point of view of the field of quantum cosmology and will be pursued in the future.

\section*{Acknowledgement}
The author would like to thank Przemysław Małkiewicz, Nils A. Nilsson, Tim Schmitz and Martin Bojowald for providing useful comments and references during preparation of this paper.\\
The work was supported by Narodowe Centrum Nauki with Decision No. DEC-2017/27/N/ST2/01964.

\appendix
\section{Phase space of fiducial vectors}\label{Ap:A}
The computations below present an explicit phase space construction for both `classical' and `quantum' degrees of freedom and might be useful for running the basis truncation method from section \ref{sec:approximate}. For these calculations a specific basis is needed, the case of a discrete (finite or countable) set of a basis vectors is chosen and the physical centering conditions are imposed. The basis is assumed to be orthonormal and any fiducial vector can be decomposed as a linear combination
\begin{equation}
| \phi \rangle = \sum_i \lambda_i(t) | e_i \rangle,
\end{equation}
where $\lambda 's$ are time dependent coefficients and $| e_i \rangle$ are basis vectors. The Legendre transformation yields
\begin{equation}
\dot{q} \mapsto \frac{\delta S}{\delta \dot{q}} = p,\ \dot{\lambda}_i \mapsto \frac{\delta S}{\delta \dot{\lambda}_i} = i \lambda_i^*,
\end{equation}
and the canonical structure is given by the symplectic form
\begin{equation}\label{eq:Appen:symplectic}
\omega = dq \wedge dp + i d \lambda_i \wedge \lambda_i^*,
\end{equation}
where the repeated index summation is assumed.\\
The total hamiltonian reads
\begin{equation}\label{eq:Appen:Ht}
H_T = \frac{p^2}{2m}+ \frac{1}{2m}(\p^2)_{ij} \lambda^*_i \lambda_j + \frac{p}{m} \p_{ij} \lambda^*_i \lambda_j + \hat{V}_{ij}(q) \lambda^*_i \lambda_j + \alpha \q_{ij} \lambda^*_i \lambda_j + \beta \p_{ij} \lambda^*_i \lambda_j,
\end{equation}
where $\hat{O}_{ij} = \langle e_i | \hat{O} | e_j \rangle$ are matrix representations of the operators. The consistency conditions for physical centering constraints imply
\begin{equation}
\alpha = \frac{[\p,\frac{1}{2m}\p^2+ \frac{1}{m}p \p + V(q)]_{ij}\lambda^*_i \lambda_j}{[\q,\p]_{ij}\lambda^*_i \lambda_j},\ \ \beta = - \frac{[\q,\frac{1}{2m}\p^2+ \frac{1}{m}p \p + V(q)]_{ij}\lambda^*_i \lambda_j}{[\q,\p]_{ij}\lambda^*_i \lambda_j}.
\end{equation}
The commutators are deliberately not computed, as for an approximate analysis their outcomes will differ from the commutators calculated on full Hilbert space.\\
Making use of the symplectic form (\ref{eq:Appen:symplectic}) and the Hamiltonian (\ref{eq:Appen:Ht}) it is straightforward to derive the Hamilton equations. The dynamical laws for the position $q$ and the momentum $p$ read
\begin{subequations}
\begin{align}
\dot{q} &\approx 2p,\\
\dot{p} &\approx - \frac{\partial V_{ij}(q)}{\partial q} \lambda^*_i \lambda_j
\end{align}
\end{subequations}
The above equations govern the dynamics of `classical' degrees of freedom in the system. The first equation can be immediately recognized as the definition of momentum which is identical to its classical counterpart. The second equation includes the coupling to the quantum degrees of freedom through $\lambda$'s.\\
The dynamics of the quantum variables, $\lambda$’s, reads
\begin{equation}
i \dot{\lambda}_i = \left(\frac{1}{2m}(\p^2)_{ij}+ \frac{1}{m}p \p_{ij} + V_{ij}(q) \right) \lambda_j,
\end{equation}
where the dynamics of the conjugate variable $\lambda_i^*$ is obtained by taking the hermitian conjugate of the above equation.
The above equations govern the dynamics of purely quantum degrees of freedom associated with the shape of the wave-function and are absent in the classical mechanics. The dynamics of $\lambda$’s is coupled to `classical' degrees of freedom through the position, $q$, which features in the potential $V_{ij} (q)$ and in the coefficients $\alpha$ and $\beta$.

\printbibliography

\end{document}